\documentclass[12pt]{article}

\usepackage{amsmath,amssymb}
\usepackage{amsfonts}
\usepackage{graphicx}
\usepackage{a4wide}

\begin{document}


\begin{center} {\bf M. V. Burnashev, H. Yamamoto} \end{center}

\vskip 0.4cm

\begin{center}
{\large\bf ON ZERO-RATE ERROR EXPONENT \\ FOR BSC WITH NOISY
FEEDBACK \footnote[1]{The research described in this publication
was made possible in part by the Russian Fund for Fundamental
Research (project number 06-01-00226).}}
\end{center}

{\begin{quotation} \normalsize For the information transmission a
binary symmetric channel is used. There is also another noisy
binary symmetric channel (feedback channel), and the transmitter
observes without delay all the outputs of the forward channel via
that feedback channel. The transmission of a nonexponential number
of messages (i.e. the transmission rate equals zero) is considered.
The achievable decoding error exponent for such a combination
of channels is investigated. It is shown that if the crossover
probability of the feedback channel is less than a certain
positive value, then the achievable error exponent is better than
the similar error exponent of the no-feedback channel.

The transmission method described and the corresponding lower bound
for the error exponent can be strengthened, and also extended to the
positive \\ transmission rates.
\end{quotation}}

\vskip 0.7cm

\begin{center}
{\large\bf \S\,1. Introduction and main results}
\end{center}

The binary symmetric channel ${\rm BSC}(p)$ with crossover
probability $0 < p < 1/2$ (and $q = 1-p$) is considered. It is
assumed that there is the feedback ${\rm BSC}(p_{1})$ channel, and
the transmitter observes (without delay) all outputs of the forward
${\rm BSC}(p)$ channel via that noisy feedback channel. No coding
is used in the feedback channel (i.e. the receiver simply
re-transmits all received outputs to the transmitter). In words,
the feedback channel is ``passive''.

Since the Shannon's paper \cite{Shannon56} it has been known that
even the noiseless feedback does not increase the capacity of the
${\rm BSC}$ (or any other memoryless channel). However, the
feedback can improve the decoding error probability (or simplify the
effective transmission method). In the case of ${\rm BSC}$ with
noiseless feedback investigations of the decoding error probability
(or its best error exponent - {\it channel reliability function})
have been actively studied since Dobrushin \cite{Dob},
Horstein \cite{Hors1} and Berlekamp \cite{Ber1}. Some
characteristics of a number of efficient transmission methods have
been investigated (see, for example, [1--10]). Generally, the case
of ${\rm BSC}$ with noiseless feedback is reasonably well
investigated (although there are still some important open
problems).

The case of noisy feedback was not investigated. It was not even
known whether such feedback can improve the error exponent of the
no-feedback case. In this respect, only two recent papers
\cite{DrapSah1, KimLapW} can probably be mentioned, but both of
them consider different problems. In the paper \cite{DrapSah1}
the {\it variable-length} coding (i.e. non-block codes) is used
under a different error criterion. Moreover, it is assumed that at
certain moments an error-free mechanism in the feedback is
available. In the paper \cite{KimLapW} Gaussian channel with only
the {\it average power} constraint is considered. Such constraint
allows using some methods which are unavailable in the
case of discrete channels.

We try to explain the reason why the noisy feedback case is so badly
investigated, and what creates the main difficulty (how we see it).
In the noiseless feedback case the transmitter
{\it at any moment} may change its coding function (transmission
method), and the receiver will know exactly about this change. Such
an ideal mutual understanding (mutual coordination) between the
transmitter and the receiver was very important for all results on
the noiseless feedback case [1--10]. If we try to apply any of the
transmission methods from [1--10] to a noisy feedback case, we
find that the transmitter and the receiver rather quickly loose
their mutual coordination. Due to noise in the feedback link
they can achieve mutual coordination only in some probabilistic
sense. In particular, if the transmitter wants to change its coding
function at some moment $t$, it should know with high reliability
the current output values of some functions (e.g. posterior message
probabilities) at the receiver. Of course, it takes a certain time
to achieve high reliability of such knowledge. For that reason, the
transmitter should probably change the coding function not very
often (i.e. only after accumulating some very reliable information
on the receiver uncertainty).

The following geometrical picture explains that description. Let
${\cal D}_{1},\ldots,{\cal D}_{M}$ be the \\ optimal decoding
regions of messages $\theta_{1},\ldots,\theta_{M}$, respectively.
The boundary part of each region  ${\cal D}_{i}$ gives the main
contribution to the decoding error. The transmitter aim is to
``push'' the output into the corresponding region ${\cal D}_{i}$.
The best transmitter strategy is to ``push'' the current output in
the direction ``orthogonal'' to the closest boundary of the true
region ${\cal D}_{i}$. Then, essentially, two cases are possible.

1) If all ${\cal D}_{i}$ are ``round-shaped'' (i.e. similar to
``balls''), then they have the centers, and therefore the best transmitting
strategy is to send the center of the corresponding ``ball'' (and
that strategy does not depend on the output signals). It
automatically pushes the output in the direction ``orthogonal'' to
the closest boundary. This situation takes place for sufficiently
high transmission rates $R$. Then, even noiseless feedback cannot
improve the error exponent.

2) The situation becomes quite different if the optimal decoding
regions $\{{\cal D}_{i}\}$ are not ``round-shaped'' (and so, they do
not have the natural centers). Now the best transmitter strategy
depends on the current output location. For the case of three
messages, it is depicted in Fig. 1. Let the message $\theta_{1}$ be
transmitted, and then the transmitter pushes the output into the
region ${\cal D}_{1}$. If the current output is close to the point
$A$ (i.e. to two other possible regions), then best is to push the
output simultaneously away from both competitive regions. On the
contrary, if the current output is close to the point $B$ (i.e. it
is much closer to the competitive region ${\cal D}_{2}$ than to
${\cal D}_{3}$), then best is to push the output mainly away from
the region ${\cal D}_{2}$, paying less attention to the other
region ${\cal D}_{3}$.

That best strategy is possible only if the transmitter knows exactly
the current output location (i.e. if there is noiseless feedback).
If there is no any feedback then the transmitter knows nothing on
the current output location, and there is no sense to change the
push direction. The situation becomes ``fuzzy'', if the transmitter
knows only approximately the current output location (i.e. if there
is noisy feedback).

In this paper we realize those arguments, allowing only one fixed
time moment when the transmitter may change the coding function. At
that moment the transmitter, using observations over the feedback
channel, finds two messages which are the most probable for the
receiver. After that the transmitter only helps the receiver to
decide between those two messages. Of course, an error is possible
when choosing those two most probable messages. However, we show
that if the crossover probability of the feedback channel is less
than the certain positive value, then the probability of making an
error in that choice is sufficiently small. Such simple transmission
method (together with the properly chosen decoding) allows already
to improve the decoding error probability in comparison with the
no-feedback case.

Of course, if the feedback channel noise is rather small then it is
possible to use a larger number of such ``switching'' moments, and
to improve further the error probability exponent. In the limit (if
the feedback channel noise is very small), using a growing number
of switching moments, we can achieve the noiseless feedback case
performance.

We consider the case when the overall transmission time $n$ and
$M = M_{n}$ equiprobable messages $\{\theta_{1},\ldots,\theta_{M}\}$
are given. It is assumed that $M_{n} \to \infty$, but
$\ln M_{n} = o(n)$ as $n \to \infty$, i.e. the transmission rate
$R = 0$. After the moment $n$, the receiver makes a decision
${\hat \theta}$ on the message transmitted. We limit ourselves here
only to the case $R = 0$, since in that case the difficulties of
using noisy feedback are seen most clearly. In the case of a
positive transmission rate $R$ (it will be considered in another
publication) some additional technical difficulties appear, which we
want to avoid for a while. It should also be mentioned that the
investigation of the best error exponent for $R = 0$ even for the
noiseless feedback case is not a simple task \cite{Ber1}.

As a result, we show that if the crossover probability $p_{1}$
of the feedback channel ${\rm BSC}(p_{1})$ is less then the certain
positive value $p_{0}(p)$, then it is possible to improve the best
error exponent $E(p)$ of ${\rm BSC}(p)$ without feedback. The
transmission method with one ``switching'' moment, giving such an
improvement, is described in  \S\,3.

Denote by $E(p)$ the best error exponent for $M_{n}$ codewords
over ${\rm BSC}(p)$ without feedback, i.e.
\begin{equation}\label{deferexp}
E(p) = \limsup_{n \to \infty}\,\frac{1}{n}\,
\ln \frac{1}{P_{\rm e}(M_{n},n,p)}\,, \qquad \ln M_{n} = o(n)\,,
\end{equation}
where $P_{\rm e}(M_{n},n,p)$ is the minimal possible decoding error
probability $P_{\rm e}$ for all codes of length $n$. Clearly, we
have
\begin{equation}\label{E3}
\begin{gathered}
E(p) = \frac{1}{4}\ln \frac{1}{4pq}\,.
\end{gathered}
\end{equation}
Indeed, the minimal Hamming distance of any such code does not
exceed $n/2$ (Plotkin bound). On the other hand, due to the
Varshamov-Gilbert bound there exist codes with approximately such
minimal distance. If $E(R,p)$ -- the reliability function of the
${\rm BSC}(p)$ without feedback, then $E(p) = E(0,p)$.

Denote by $E_{2}(p)$ the best error exponent for two codewords over
${\rm BSC}(p)$ (it remains the same for the channel with noiseless
feedback, as well). Clearly, we have
$$
\begin{gathered}
E_{2}(p) = \frac{1}{2}\ln \frac{1}{4pq}\,.
\end{gathered}
$$

Denote by $F(p)$ the best error exponent for $M_{n}$ messages
over ${\rm BSC}(p)$ with noiseless feedback. It is defined
similarly to (\ref{deferexp}), where $P_{\rm e}(M_{n},n,p)$ is the
minimal possible decoding error probability for all transmission
methods. Denote also by $F_{3}(p)$ the best error exponent for three
messages over ${\rm BSC}(p)$ with noiseless feedback. Then
\cite{Ber1}
\begin{equation}\label{F3}
\begin{gathered}
F(p) = F_{3}(p) = -\ln\left(p^{1/3}q^{2/3} + q^{1/3}p^{2/3}\right).
\end{gathered}
\end{equation}
If $F(R,p)$ -- the reliability function of such channel, then
$F(p) = F(0,p)$.

Denote by $F(p,p_{1})$ the best error exponent for $M_{n}$ messages
transmitted over the ${\rm BSC}(p)$ with the noisy
${\rm BSC}(p_{1})$ feedback channel. Clearly, $E(p) \leq
F(p,p_{1}) \leq F(p)$ for all $p,p_{1}$. In particular,
$F(p,0) = F(p),\, F(p,1/2) = E(p)$. Moreover,
$E(p) < F(p) < E_{2}(p),\,0 < p < 1/2$. Let $r(p) = F(p)/E(p)$.
The function $r(p)$ monotonically increases on $p$, and, in
particular,
$$
r(0)= \lim\limits_{p \to 0}r(p)= 4/3\,,\quad r(0.01) \approx 1.67\,,
\quad r(1/2) = \lim\limits_{p \uparrow 1/2} = 16/9 \approx 1.78\,.
$$
More exactly, if $p=(1-\varepsilon)/2$ then ($\varepsilon \to 0$)
$$
E(p) = \varepsilon^{2}/4 + O(\varepsilon^{4}),\quad F(p) =
4\varepsilon^{2}/9 + O(\varepsilon^{4}).
$$

Below in the paper $f \sim g$ means
$n^{-1}\ln f = n^{-1}\ln g + o(1),\,n \to \infty$, and
$f \lesssim g$ means
$n^{-1}\ln f \leq n^{-1}\ln g + o(1),\,n \to \infty$.

To formulate the paper main result, introduce the functions:
\begin{equation}\label{Pen2}
\begin{gathered}
h(x) = -x\ln x - (1-x)\ln(1-x), \\
z = q/p\,, \quad z_{1} = q_{1}/p_{1}\,, \\
3G_{1}(t,p) = \ln \frac{1}{qp^{2}} -
\max_{a}\left\{2h(a) + h(a+t)+ (a+t)\ln z\right\}, \\
3G_{2}(t,p,p_{1}) = (2c_{0}+t)\ln z - h(c_{0}+t) - h(c_{0})
+ [2+t-2(1+t)b_{1}]\ln z_{1} - \\
- (1+t)h(b_{1})- (1-t)h\left[\frac{(1+t)b_{1}-t}{1-t}\right] -
2\ln (qq_{1})\,, \\
c_{0}(t,p) = \frac{2(1-t)}{2 + t(z^{2}-1) +
\sqrt{4z^{2} + t^{2}(z^{2}-1)^{2}}}\,, \\
b_{1}(t,p_{1}) = \frac{2z_{1}^{2}}{(2+t)z_{1}^{2} - t +
\sqrt{4z_{1}^{2} + (z_{1}^{2} - 1)^{2}t^{2}}}\,,
\end{gathered}
\end{equation}
The optimal
$a_{0} = a_{0}(p,t)$ in (\ref{Pen2}) is defined as the unique root
of the equation
\begin{equation}\label{opta}
q(1-a)^{2}(1-a-t) = pa^{2}(a+t)\,.
\end{equation}
We have $a_{0}(p,0) = a_{0}(p)$, where $a_{0}(p)$ is the same as
defined below in (\ref{a0}).

Introduce also the function $p_{0}(p)$ as the unique root of the
equation
\begin{equation}\label{defp0}
3G_{2}(1/2-p,p,p_{0}) = \ln \frac{1}{4pq}\,, \qquad 0 < p < 1/2\,.
\end{equation}

Denote by $F_{1}(p,p_{1})$ the error exponent for the transmission
method with one switching moment, described in \S 3. Clearly,
$F_{1}(p,p_{1}) \leq F(p,p_{1})$ for all $p,p_{1}$.
The paper main result is

\medskip

{T h e o r e m}. {\it If $p_{1} < p_{0}(p)$, then}
\begin{equation}\label{Pe00}
\begin{gathered}
F(p,p_{1}) \geq F_{1}(p,p_{1}) = \max_{t}
\frac{6\min\left\{G_{1}(t,p),G_{2}(t,p,p_{1})\right\}E(p)}
{3\min\left\{G_{1}(t,p),G_{2}(t,p,p_{1})\right\} + 4E(p)} > E(p)\,.
\end{gathered}
\end{equation}

\medskip

The function $G_{1}(t,p)$ monotonically decreases on $t$, and
$G_{1}(0,p) = F(p)$. On the other hand, the function
$G_{2}(t,p,p_{1})$ monotonically increases on $t$. Moreover,
$G_{2}(0,p,p_{1}) = 0$, and \\ $G_{2}(t,p,0) = \infty,\,t > 0$.

The function $p_{0}(p),\,0 < p < 1/2$, is positive and monotonically
increases on $p$. Its plot is shown in Fig. 2.

\medskip

{E x a m p l e\, 1}. Consider the case $p \to 0$. Then
$$
p_{0}(p) = \frac{16p}{27}\left(1 + o(1)\right) \,.
$$
The approximation $p_{0}(p) \approx p/2$ is quite accurate for
$p \leq 0.01$.

\medskip

{E x a m p l e\, 2}. Consider the opposite asymptotic case
$p=(1-\varepsilon)/2,\,\varepsilon \to 0$, and
$t \leq 1/2-p = \varepsilon/2$. Then
$a = a_{0}(p,t) = 1/2 - \rho,\,\rho \to 0$, and after standard
algebra we get
$$
\begin{gathered}
\rho = \frac{2t-\varepsilon}{6} + O(\varepsilon^{2})\,,
\end{gathered}
$$
which gives
$$
\begin{gathered}
G_{1}(t,p) = \frac{4(\varepsilon^{2}-\varepsilon t + t^{2})}{9} +
O\left(\varepsilon^{4}\right)\,, \\
G_{2}(t,p,p_{1}) = \frac{t^{2}}{12p_{1}q_{1}} +
O\left(\varepsilon^{3}\right), \qquad t\leq
\frac{\varepsilon}{2}\,.
\end{gathered}
$$
If $G_{1}(t,p) = G_{2}(t,p,p_{1})$, then
$$
t = \frac{4\varepsilon \sqrt{p_{1}q_{1}}}{\sqrt{3-12p_{1}q_{1}} +
2\sqrt{p_{1}q_{1}}} + O(\varepsilon^{2}),
$$
which gives
$$
\min\left\{G_{1}(t,p),G_{2}(t,p,p_{1})\right\} =
\frac{4\varepsilon^{2}}{3\left[\sqrt{3-12p_{1}q_{1}} +
2\sqrt{p_{1}q_{1}}\right]^{2}} + O\left(\varepsilon^{3}\right).
$$
The condition $t \leq \varepsilon/2$ is equivalent to the
inequality $16p_{1}q_{1} \leq 1$, which means that
$$
\lim_{p\to 1/2}p_{0}(p) = \frac{1}{4(2+\sqrt{3})} \approx
\frac{1}{14.93} \approx 0.067\,.
$$
For $p_{1} \to 0$ we get
\begin{equation}\label{asPeM}
\begin{gathered}
F(p,p_{1}) \geq \frac{8E(p)}{7}\left[1 -
\frac{4\sqrt{3p_{1}}}{7} + \frac{104p_{1}}{49} +
O\left(p_{1}^{3/2}\right)\right].
\end{gathered}
\end{equation}
In words, for small $p_{1}$ the strategy described in \S 3 gives $14\%$
gain over the no-feedback channel.

\medskip

{C o r o l l a r y}. {\it If $p_{1} = 0$, then}
\begin{equation}\label{Pef=0}
\begin{gathered}
F_{1}(p,0) = \frac{6E(p)F(p)}{4E(p) + 3F(p)} > E(p)\,, \qquad
0 < p < 1/2\,.
\end{gathered}
\end{equation}

\medskip

{E x a m p l e\, 3}. We have $F_{1}(p,p_{1}) \to F_{1}(p,0)$ as
$p_{1} \to 0$. We investigate the rate of that convergence since it
gives some idea on when the noisy feedback behaves like the
noiseless feedback. If $p_{1} \to 0$, then the optimal $t \to 0$.
For a fixed $0 < p < 1/2$ and $t \to 0$ for the root $a(t,p)$ of the
equation (\ref{opta}) we have
$$
a(t,p) = a_{0}(p) - \frac{t}{3} + O(t^{2})\,,
$$
which gives
$$
G_{1}(t,p) = F(p) - \frac{2t}{9}\ln z + O(t^{2})\,.
$$
We also can get as $p_{1},t \to 0$
$$
\begin{gathered}
c_{0}(t,p) = p- \frac{t}{2}+ \frac{(q-p)t^{2}}{8qp} + O(t^{3})\,, \\
b_{1}(t,p_{1}) = 1 - t + O\left(\frac{p_{1}^{2}}{p_{1}+ t}\right) +
O\left(t^{2}\right),
\end{gathered}
$$
which gives
$$
\begin{gathered}
3G_{2}(t,p,p_{1}) = -t\ln p_{1} + O\left(p_{1}\ln p_{1}\right) +
O\left(t\ln t\right) + O\left(t^{2}\ln p_{1}\right).
\end{gathered}
$$
If $G_{1}(t,p) = G_{2}(t,p,p_{1})$, then
$$
t = \frac{3F(p)}{\ln(1/p_{1})}\left[1+o(1)\right],
$$
and
$$
\min\left\{G_{1}(t,p),G_{2}(t,p,p_{1})\right\} = F(p)\left[1 -
\frac{2\ln z}{3\ln(1/p_{1})} + o\left(\frac{1}{\ln(1/p_{1})}\right)
\right].
$$
As a result, we get as $p_{1} \to 0$
$$
\begin{gathered}
F_{1}(p,p_{1}) = F_{1}(p,0)\left[1 -\frac{8E\ln z}
{3(4E + 3F)\ln(1/p_{1})} + o\left(\frac{1}{\ln(1/p_{1})}\right)
\right].
\end{gathered}
$$

\medskip

{\it Remark} 1. The transmission method described in \S\,3, reduces
the problem to testing of two most probable messages (at the fixed
moment). Such strategy is not optimal even for one switching
moment. But it is relatively simple for investigation, and it gives
already a reasonable improvement over the no-feedback case.

In \S\,2 the transmission method with one switching moment for
the channel with noiseless feedback is described and investigated.
In particular, the formula (\ref{Pef=0}) is proved. In \S\,3
that transmission method (slightly modified) is investigated for
the channel with noisy feedback, and the theorem is proved.
In \S\,4 the simple transmission method with active feedback is
considered.

The preliminary (and simplified) paper version (without detailed
proofs) for $M = 3$ messages was published as \cite{BurYam1}.

\medskip


\begin{center}
{\large\bf \S\,2. Channel with noiseless feedback. Proof of the
formula (\ref{Pef=0}).}
\end{center}

We start with the noiseless feedback case and describe the
transmission method which will be used for noisy feedback as well.
Moreover, in the noisy feedback case we will need some formulas
from that case.

Consider the ${\rm BSC}(p)$ with noiseless feedback and $M$ messages
$\theta_{1},\ldots,\theta_{M}$. We assume that $M_{n} \to \infty$,
but $\ln M_{n} = o(n)$ as $n \to \infty$. We set some
$\gamma \in [0,1]$ (it will be chosen later) and divide the
total transmission period $[0,n]$ on two phases: $[0,\gamma n]$
(phase I) and $(\gamma n, n]$ (phase II). We perform as follows:

1) On phase I (i.e. on $[0,\gamma n]$) we use a code of $M$
codewords $\{\mbox{\boldmath $x$}_{i}\}$ such that
$d\left(\mbox{\boldmath $x$}_{i},\mbox{\boldmath $x$}_{j}\right) =
\gamma n/2 + o(n),\;i \neq j$ (existence of such ``almost'' a
simplex code can be shown using random choice of codewords). On that
phase the transmitter only observes via the feedback channel
outputs of the forward channel, but does not change the
transmission method.

2) Let $\mbox{\boldmath $x$}$ be the transmitted codeword (of length
$\gamma n$) and  $\mbox{\boldmath $y$}$ be the received (by the
receiver) block. After phase I, based on the block
$\mbox{\boldmath $y$}$, the transmitter selects two messages
$\theta_{i},\theta_{j}$ (codewords
$\mbox{\boldmath $x$}_{i},\mbox{\boldmath $x$}_{j}$) which are the
most probable for the receiver, and ignore all the remaining
messages $\{\theta_{k}\}$. Then, on phase II (i.e. on
$(\gamma n, n]$) the transmitter helps the receiver only to decide
between those two most probable messages $\theta_{i},\theta_{j}$,
using two opposite codewords of length $(1-\gamma) n$. After moment
$n$ the receiver makes a decision between those two remaining
messages $\theta_{i},\theta_{j}$ (based on all received on $[0,n]$
signals).

Clearly, a decoding error occurs in the following two cases.

1) After phase I the true message is not among two most probable
messages. We denote that probability $P_{1}$.

2) After phase I the true message is among two most probable, but
after phase II the true message is not the most probable. We denote
that probability $P_{2}$.

Then for the total decoding error probability $P_{\rm e}$ we have
\begin{equation}\label{Petotal}
P_{\rm e} \leq P_{1} + P_{2}\,.
\end{equation}

To evaluate the probabilities $P_{1}$ and $P_{2}$, without loss of
generality, we assume that the message ${\bf \theta}_{1}$ is
transmitted. We start with the probability $P_{1}$. Denote
$d\left(\mbox{\boldmath $x$},\mbox{\boldmath $y$}\right)$ the
Hamming distance between $\mbox{\boldmath $x$}$ and
$\mbox{\boldmath $y$}$, and
$d_{i} = d\left(\mbox{\boldmath $x$}_{i},\mbox{\boldmath $y$}
\right)$. Then
\begin{equation}\label{P1up1}
P_{1} \leq \sum_{i > j > 1}{\bf P}\{d_{1} \geq \max\{d_{i},d_{j}\}|
\mbox{\boldmath $x$}_{1}\}.
\end{equation}

We use the following auxiliary result (see proof in Appendix).

\medskip

{L e m m a}. 1) {\it Let $\mbox{\boldmath $x$}_{1},
\mbox{\boldmath $x$}_{2},\mbox{\boldmath $x$}_{3}$ be the codewords
of length $m$. Denote $d_{ij} = d\left(\mbox{\boldmath $x$}_{i},
\mbox{\boldmath $x$}_{j}\right),\,\\
d_{i} = d\left(\mbox{\boldmath $x$}_{i},\mbox{\boldmath $y$}
\right)$. Assuming that
$d_{12} = d_{13} = d_{23} = 2m/3 + o(m),\,m \to \infty$, consider
the probability
$$
P_{1}(t,t_{1}) = {\bf P}\left(d_{2} = d_{1} +  \dfrac{2tm}{3}+ o(m);
d_{3}=  d_{1}+ \dfrac{2t_{1}m}{3}+ o(m)\Big|\mbox{\boldmath $x$}_{1}
\right).
$$
Then
\begin{equation}\label{lem4}
\begin{gathered}
\frac{3}{m}\ln P_{1}(t,t_{1}) = \ln(p^{2}q) + f(t,t_{1}) + o(1)\,,
\qquad |t| \leq 1, \quad |t_{1}| \leq 1 \,,
\end{gathered}
\end{equation}
where
\begin{equation}\label{lem4b}
\begin{gathered}
f(t,t_{1}) = \max_{a}f(a,t,t_{1}) = f(a_{0},t,t_{1}), \\
f(a,t,t_{1}) = h(a)+ h(a+t) + h(a+t_{1}) + (a+t_{1}+t)\ln z \,,
\end{gathered}
\end{equation}
and $a_{0} = a_{0}(t,t_{1})$ is the unique root of the equation
$$
f'_{a} = \ln\frac{1-a}{a} + \ln\frac{1-a+t}{a-t} +
\ln\frac{1-a-t_{1}}{a+t_{1}} + \ln z = 0\,.
$$
The function $f(t,t_{1})$ monotone increases on
$t_{1} \leq (1-2p+t)/2$, and monotone decreases on
$t_{1} \geq (1-2p+t)/2$.}

2) {\it For any $|t| \leq 1$ and $t_{1} \leq (1-2p+t)/2$, we have }
\begin{equation}\label{lem4a}
\begin{gathered}
{\bf P}\left(d_{2} \leq d_{1} +  \dfrac{2tm}{3} ;
d_{3} \leq d_{1}+ \dfrac{2t_{1}m}{3}\Big|\mbox{\boldmath $x$}_{1}
\right) = P_{1}(t,t_{1})e^{o(m)}\,, \qquad m \to \infty\,.
\end{gathered}
\end{equation}

\medskip

Note that the number of summation terms in the right-hand side of
(\ref{P1up1}) does not exceed $M^{2} = e^{o(n)}$. Any three
codewords $\mbox{\boldmath $x$}_{1},\mbox{\boldmath $x$}_{i},
\mbox{\boldmath $x$}_{j}$ have the effective length
$m = 3\gamma n/4 + o(n)$ (on the remaining $\gamma n/4+ o(n)$
positions they have equal coordinates) and mutual distances
$d\left(\mbox{\boldmath $x$}_{k},\mbox{\boldmath $x$}_{l}\right) =
2m/3 + o(m),\,k \neq l$. Then using the formulas (\ref{lem4b}) and
(\ref{lem4a}) with $t = t_{1} = 0$, we have
$$
\begin{gathered}
m^{-1}\ln P_{1} = \frac{1}{3}\ln (p^{2}q) +
\frac{1}{3}\max_{a}\left\{3h(a) + a\ln z\right\} + o(1) = \\
= \ln \left(p^{1/3}q^{2/3} + p^{2/3}q^{1/3}\right) = - F(p) +o(1)\,,
\end{gathered}
$$
where $F(p)$ is defined in (\ref{F3}), and the optimal $a = a_{0}$
is given by
\begin{equation}\label{a0}
a_{0} = a_{0}(p) = \frac{q^{1/3}}{p^{1/3}+q^{1/3}}\,.
\end{equation}
As a result, from (\ref{P1up1}) we get
\begin{equation}\label{Pe12}
\ln \frac{1}{P_{1}} = \frac{3}{4}\gamma F(p)n + o(n)\,,
\qquad 0 \leq p \leq 1/2\,.
\end{equation}

{\it Remark } 2. Let $\{\mbox{\boldmath $x$}_{1},
\mbox{\boldmath $x$}_{2},\mbox{\boldmath $x$}_{3}\}$ be a
simplex code of length $n$. Then
\begin{equation}\label{opt3}
-\frac{1}{n}\ln {\bf P}\{d\left(\mbox{\boldmath $x$}_{1},
\mbox{\boldmath $y$}\right) \geq \max
\{d\left(\mbox{\boldmath $x$}_{2},
\mbox{\boldmath $y$}\right),d\left(\mbox{\boldmath $x$}_{3},
\mbox{\boldmath $y$}\right)\}\big|\mbox{\boldmath $x$}_{1}\} =
F_{3}(p) + o(1)\,, \qquad n \to \infty\,.
\end{equation}
It explains the meaning of the value $F(p) = F_{3}(p)$.

Now we evaluate the probability $P_{2}$. On phase I (of length
$\gamma n$) all the distances among codewords are equal to
$\gamma n/2 + o(n)$. On phase II (of length $(1-\gamma)n$)
the distance between two remaining codewords equals $(1-\gamma)n$.
Therefore the total distance between the true and any concurrent
codeword equals $(1-\gamma/2)n$. Therefore
$$
P_{2} \leq M{\bf P}\{\mbox{error when testing two codewords on
distance $(1-\gamma/2)n$}\},
$$
and then
\begin{equation}\label{Pe2}
\begin{gathered}
\frac{1}{n}\ln P_{2} = \frac{(1-\gamma/2)}{2}\ln(4pq) + o(1) =
-(2-\gamma)E(p) + o(1)\,.
\end{gathered}
\end{equation}

As a result, from (\ref{Petotal}), (\ref{Pe12}) and (\ref{Pe2})
for the decoding error probability $P_{\rm e}$ we have
$$
\frac{1}{n}\ln P_{\rm e} \leq \frac{1}{n}
\max\left\{\ln P_{1}, \ln P_{2}\right\} \leq -\min
\left\{\frac{3}{4}\gamma F(p),(2-\gamma)E(p)\right\} + o(1)\,.
$$
We choose $\gamma = \gamma_{0}$ such that $P_{1} = P_{2}$, i.e. set
\begin{equation*}
\begin{gathered}
\gamma_{0} = \frac{8E(p)}{4E(p) + 3F(p)}\,,
\end{gathered}
\end{equation*}
and then for $0 < p < 1/2$ get the formula (\ref{Pef=0}).

If $p=(1-\varepsilon)/2,\,\varepsilon \to 0$, then
$$
F(p,0) \to \frac{8}{7}E(p)\,, \qquad p \to 1/2\,,
$$
i.e. such strategy with one switching moment gives $14\%$ gain over
the no-feedback case (the best strategy without limit on the number
of switching moments gives $78\%$ gain).

\medskip

\newpage
\begin{center}
{\large\bf \S\,3.  Channel with noisy feedback. Proof of theorem}
\end{center}

In the noisy feedback case, still using one switching moment, we
will slightly modify the transmission method from \S\,2
(especially, its decoding method).

{\bf Transmission}. Again we set a number $0 < \gamma < 1$. On phase
I, of length $\gamma n$, we use an ``almost'' simplex code. Let
$\mbox{\boldmath $x$}$ be the transmitted codeword (of length
$\gamma n$), $\mbox{\boldmath $y$}$ be the received (by the
receiver) block, and $\mbox{\boldmath $x$}'$ be the received (by the
transmitter) block. Based on the transmitted codeword
$\mbox{\boldmath $x$}$ and the received block
$\mbox{\boldmath $x$}'$, the transmitter selects two messages
$\theta_{i},\theta_{j}$ which look most probable for the receiver.

If the true message is among those two selected messages
$\theta_{i},\theta_{j}$, then, on phase II (i.e. on $(\gamma n, n]$)
the transmitter uses the two opposite codewords of length
$(1-\gamma) n$ to help the receiver to decide between those
two most probable messages. For example, the transmitter uses
all-zeros and all-ones codewords.

If the true message is not among two selected messages
$\theta_{i},\theta_{j}$, then, on phase II the transmitter sends
an intermediate block (say, half-zeros and half-ones). In any case,
such event will be treated as an error.

{\bf Decoding}. We set an additional  number $t > 0$. Arrange the
distances $\{d(\mbox{\boldmath $x$}_{i},\mbox{\boldmath $y$}),\,
i=1,\ldots,M\}$ in the increasing order, denoting
$$
d^{(1)} = \min_{i} d(\mbox{\boldmath $x$}_{i},\mbox{\boldmath $y$})
\leq  d^{(2)} \leq \ldots \leq d^{(M)} =
\max_{i} d(\mbox{\boldmath $x$}_{i},\mbox{\boldmath $y$}),
$$
(in case of tie we use any order). Let also
$\mbox{\boldmath $x$}^{1},\ldots,\mbox{\boldmath $x$}^{M}$ be the
ranking of codewords after phase I, i.e $\mbox{\boldmath $x$}^{1}$
is the most probable codeword, etc. There are possible two cases.

C a s e \,1. If $d^{(3)} \leq d^{(2)} + t\gamma n/2$, then the
receiver makes the decoding immediately after phase I (in favor
of the closest to $\mbox{\boldmath $y$}$ codeword). Although the
transmitter still continues transmission, the receiver has already
made its decision.

C a s e \,2. If $d^{(3)} > d^{(2)} + t\gamma n/2$, then after
phase I the receiver selects two most probable messages
$\theta_{i},\theta_{j}$, and after transmission on phase II
(i.e. after moment $n$) makes a decision between those two remaining
messages $\theta_{i},\theta_{j}$ in favor of more probable of them.

In order to perform in agreement with the receiver, in the case 2
it is important that the transmitter can correctly identify two
messages $\theta_{i},\theta_{j}$ which are most probable for the
receiver. Of course, an error in such selection is possible, but its
probability should be sufficiently small (which will be secured
below).

{\it Remark } 3. We separate the case 1 since after phase I, with
relatively high probability the second $\mbox{\boldmath $x$}^{2}$
and the third $\mbox{\boldmath $x$}^{3}$ ranked codewords will be
approximately equiprobable, and then it will be difficult to the
transmitter to rank them correctly. But in that case (with high
probability) the first message $\mbox{\boldmath $x$}^{1}$ will be
much more probable than $\mbox{\boldmath $x$}^{2}$ and
$\mbox{\boldmath $x$}^{3}$.

To evaluate the decoding error probability $P_{\rm e}$, denote
$P_{1}$ and $P_{2}$ the decoding error probability in the case 1
(i.e. after phase I), and in the case 2 (i.e. after the moment $n$)
for the noiseless feedback channel, respectively. Similarly, denote
$P_{2n}$ the decoding error probability in the case 2 for the noisy
feedback case. Then for the decoding error probability $P_{\rm e}$
we have
\begin{equation}\label{PetotaMl2}
P_{\rm e} \leq P_{1} + P_{2} + P_{2n}\,.
\end{equation}

We evaluate the probabilities $P_{1}, P_{2}, P_{2n}$ in the
right-hand side of (\ref{PetotaMl2}). For $P_{1}$ we have
\begin{equation}\label{PeII}
\begin{gathered}
P_{1} \leq M^{2}\left(P_{11} + P_{12}\right)\,,
\end{gathered}
\end{equation}
where
\begin{equation}\label{PeIIa}
\begin{gathered}
P_{11} = {\bf P}(d_{2} \leq d_{1} \leq d_{3} \leq d_{1}+
t\gamma n/2|\mbox{\boldmath $x$}_{1}), \\
P_{12} = {\bf P}\left(d_{1} \geq \max\{d_{2},d_{3}\}|
\mbox{\boldmath $x$}_{1}\right)
\end{gathered}
\end{equation}
and $d_{i} = d(\mbox{\boldmath $x$}_{i},\mbox{\boldmath $y$}),\,
i=1,\ldots,M$.

The value $P_{12}$ was already estimated in (\ref{Pe12})
(denoted there $P_{1}$). The main  contribution to $P_{1}$ is
given by the value $P_{11}$. To evaluate $P_{11}$ it is sufficient
to consider the case when the codewords
$\mbox{\boldmath $x$}_{1},\mbox{\boldmath $x$}_{2},
\mbox{\boldmath $x$}_{3}$ have length $m = 3\gamma n/4$ (on the
remaining $\gamma n/4$ positions they have equal coordinates) and
mutual distances $d\left(\mbox{\boldmath $x$}_{i},
\mbox{\boldmath $x$}_{j}\right) = 2m/3,\,i \neq j$. Then from
(\ref{lem4a}) we have
\begin{equation}\label{P11up}
\begin{gathered}
P_{11} \leq {\bf P}\left(d_{2} \leq d_{1}; d_{3} \leq d_{1} +
\dfrac{2tm}{3}\Big|\mbox{\boldmath $x$}_{1}\right)e^{o(n)} =
P_{1}(0,t)e^{o(n)}.
\end{gathered}
\end{equation}
For the value $P_{1}(0,t)$ we get from (\ref{lem4a}) and
(\ref{lem4b})
\begin{equation}\label{defPe11}
\begin{gathered}
\frac{1}{m}\ln \frac{1}{P_{1}(0,t)} = G_{1}(t,p) + o(1)\,,
\qquad t \leq \frac{1}{2} - p\,,
\end{gathered}
\end{equation}
where $G_{1}(t,p)$ is defined in (\ref{Pen2}). Moreover,
\begin{equation}\label{Pe1(1)}
\frac{1}{m}\ln \frac{1}{P_{1}(0,t)} =
\frac{1}{3}\ln \frac{1}{4pq}  + o(1) = \frac{4}{3}E(p) + o(1)\,,
\qquad t \geq \frac{1}{2} -p\,.
\end{equation}

\medskip

The function $G_{1}(t,p)$ monotonically decreases on
$t \leq 1/2-p$. Moreover, $G_{1}(0,p) = F(p)$. For $t \geq 1/2 -p$
the value $P_{1}(0,t)$ is essentially defined only by the event
$\{d_{1} \geq d_{2}\}$.

Since $P_{12} \lesssim P_{11}$, we get from (\ref{PeII}),
(\ref{Pe12}) and (\ref{defPe11})
\begin{equation}\label{Pe11a}
\begin{gathered}
\frac{4}{3\gamma n}\ln \frac{1}{P_{1}} = G_{1}(t,p) + o(1)\,,
\qquad t \leq \frac{1}{2} - p\,.
\end{gathered}
\end{equation}

For the value $P_{2}$ the formula (\ref{Pe2}) remains valid.

It remains us to evaluate $P_{2n}$, which is the probability that
the true codeword $\mbox{\boldmath $x$}_{1}$ is among two most
probable codewords for the receiver, but it is not such one for
the transmitter. Introduce the random event
\begin{equation}\label{setA11}
{\cal A} = \left\{ \begin{array}{c}
 d(\mbox{\boldmath $x$}_{3},\mbox{\boldmath $y$}) >
 \max\{d(\mbox{\boldmath $x$}_{1},\mbox{\boldmath $y$}),
d(\mbox{\boldmath $x$}_{2},\mbox{\boldmath $y$})\} + t\gamma n/2 ;\\
 d(\mbox{\boldmath $x$}_{3},\mbox{\boldmath $x$}') \leq
 \max\{d(\mbox{\boldmath $x$}_{1},\mbox{\boldmath $x$}'),
d(\mbox{\boldmath $x$}_{2},\mbox{\boldmath $x$}')\}
\end{array}
\right\}.
\end{equation}
Then
$$
P_{2n} \leq M^{2}{\mathbf P}({\cal A}|\mbox{\boldmath $x$}_{1})
e^{o(n)}.
$$
To evaluate ${\mathbf P}({\cal A}|\mbox{\boldmath $x$}_{1})$
it is convenient to use two related random events
\begin{equation}\label{setA1}
\begin{gathered}
{\cal A}_{1} = \left\{\begin{array}{c}
d(\mbox{\boldmath $x$}_{3},\mbox{\boldmath $y$}) \geq
d(\mbox{\boldmath $x$}_{2},\mbox{\boldmath $y$}) + t\gamma n/2 ; \\
d(\mbox{\boldmath $x$}_{3},\mbox{\boldmath $x$}') \leq
d(\mbox{\boldmath $x$}_{2},\mbox{\boldmath $x$}')
\end{array}
\right\}, \\
{\cal A}_{2} = \left\{\begin{array}{c}
d(\mbox{\boldmath $x$}_{3},\mbox{\boldmath $y$}) \geq
d(\mbox{\boldmath $x$}_{2},\mbox{\boldmath $y$})+ t\gamma n/2 ; \\
d(\mbox{\boldmath $x$}_{2},\mbox{\boldmath $x$}') \leq
d(\mbox{\boldmath $x$}_{3},\mbox{\boldmath $x$}') \leq
d(\mbox{\boldmath $x$}_{1},\mbox{\boldmath $x$}')
\end{array}
\right\}.
\end{gathered}
\end{equation}
Since ${\cal A} \subseteq {\cal A}_{1}\bigcup {\cal A}_{2}$,
we have
\begin{equation}\label{P2e}
{\mathbf P}({\cal A}|\mbox{\boldmath $x$}_{1})
\leq {\mathbf P}({\cal A}_{1}|\mbox{\boldmath $x$}_{1}) +
{\mathbf P}({\cal A}_{2}|\mbox{\boldmath $x$}_{1}).
\end{equation}
We may assume that the codewords
$\mbox{\boldmath $x$}_{1},\mbox{\boldmath $x$}_{2},
\mbox{\boldmath $x$}_{3}$ have length $m = 3\gamma n/4$
(on the remaining $\gamma n/4$ positions they have equal
coordinates) and mutual distances
$d\left(\mbox{\boldmath $x$}_{i},\mbox{\boldmath $x$}_{j}\right) =
2m/3,\,i \neq j$. All blocks
$\mbox{\boldmath $x$}_{1},\mbox{\boldmath $x$}_{2},
\mbox{\boldmath $x$}_{3},\mbox{\boldmath $y$},\mbox{\boldmath $x$}'$
are shown in Fig. 3, where
$a,b,c,a_{1},a_{2},b_{1},b_{2},c_{1},c_{2}$ denote the fractions of
$1$'s in the corresponding parts of the received blocks
$\mbox{\boldmath $y$}$ and $\mbox{\boldmath $x$}'$. Then in addition
to the formulas (\ref{dist1}) (see Appendix) we have
$$
\begin{gathered}
d(\mbox{\boldmath $x$}_{1},\mbox{\boldmath $x$}') =
[aa_{1} +(1-a)a_{2}+ bb_{1}+ (1-b)b_{2} + cc_{1} +
(1-c)c_{2}]m/3\,, \\
d(\mbox{\boldmath $x$}_{2},\mbox{\boldmath $x$}') =
[a(1-a_{1})+ (1-a)(1-a_{2})+ b(1-b_{1})+(1-b)(1-b_{2}) + cc_{1} +
(1-c)c_{2}]m/3\,, \\
d(\mbox{\boldmath $x$}_{3},\mbox{\boldmath $x$}') =
[a(1-a_{1})+ (1-a)(1-a_{2})+ bb_{1}+(1-b)b_{2} + c(1-c_{1}) +
(1-c)(1-c_{2})]m/3\,, \\
d(\mbox{\boldmath $y$},\mbox{\boldmath $x$}') =
[a(1-a_{1}) + (1-a)a_{2} +b(1-b_{1})+ (1-b)b_{2}+ c(1-c_{1}) +
(1-c)c_{2}]m/3\,.
\end{gathered}
$$
We start with the probability
${\mathbf P}({\cal A}_{1}|\mbox{\boldmath $x$}_{1})$. Since
$$
\begin{gathered}
d(\mbox{\boldmath $x$}_{3},\mbox{\boldmath $y$}) \geq
d(\mbox{\boldmath $x$}_{2},\mbox{\boldmath $y$}) + t\gamma n/2
\Leftrightarrow b \geq c + t\,, \\
d(\mbox{\boldmath $x$}_{3},\mbox{\boldmath $x$}') \leq
d(\mbox{\boldmath $x$}_{2},\mbox{\boldmath $x$}')
\Leftrightarrow cc_{1} + (1-c)c_{2} \geq bb_{1} + (1-b)b_{2} \,,
\end{gathered}
$$
for ${\mathbf P}({\cal A}_{1}|\mbox{\boldmath $x$}_{1})$
we have with $z = q/p,\,z_{1} = q_{1}/p_{1}$ (omitting the parts,
where $\mbox{\boldmath $x$}_{2},\mbox{\boldmath $x$}_{3}$ coincide
on all positions)
\begin{equation}\label{eqset1}
\begin{gathered}
{\mathbf P}({\cal A}_{1}|\mbox{\boldmath $x$}_{1}) =
(qq_{1})^{2m/3}\max_{b,\ldots,c_{2}}
\left\{AB\right\}\left[1+o(1)\right]
\leq (qq_{1})^{2m/3}\max_{b,\ldots,c_{2}} A \cdot
\max_{b,\ldots,c_{2}}B\left[1+o(1)\right],
\end{gathered}
\end{equation}
where
\begin{equation}\label{eqset11}
\begin{gathered}
A = \binom{m/3}{bm/3}\binom{m/3}{cm/3}z^{-m(b+c)/3}, \\
B = \binom{bm/3}{b_{1}bm/3}\binom{(1-b)m/3}{b_{2}(1-b)m/3}
\binom{cm/3}{c_{1}cm/3}\binom{(1-c)m/3}{c_{2}(1-c)m/3}
z_{1}^{-\delta(\mbox{\boldmath $y$},\mbox{\boldmath $x$}')m/3}\,, \\
\delta(\mbox{\boldmath $y$},\mbox{\boldmath $x$}') =
b(1-b_{1})+(1-b)b_{2}+ c(1-c_{1}) + (1-c)c_{2}\,,
\end{gathered}
\end{equation}
and where maximum is taken provided
\begin{equation}\label{constn1}
\begin{gathered}
b \geq c + t\,, \\
cc_{1} + (1-c)c_{2} \geq bb_{1} + (1-b)b_{2}\,.
\end{gathered}
\end{equation}
From the definition (\ref{setA1}) of the set ${\cal A}_{1}$ it is
clear that maximum of $\{AB\}$ in (\ref{eqset1}) is attained when
there are equalities in both relations (\ref{constn1}). Moreover,
there is no loss when we maximize the values $A,B$ separately. Then
we have
\begin{equation}\label{defA2)}
\begin{gathered}
3m^{-1}\ln {\mathbf P}({\cal A}_{1}|\mbox{\boldmath $x$}_{1}) \leq
2\ln (qq_{1}) + \max f + \max g + o(1)\,,
\end{gathered}
\end{equation}
where
\begin{equation}\label{defA21)}
\begin{gathered}
f = 3m^{-1}\ln A = h(b) + h(c)- (b+c)\ln z\,, \\
g = 3m^{-1}\ln B = bh(b_{1}) + (1-b)h(b_{2}) + ch(c_{1})
+ (1-c)h(c_{2})- \delta(\mbox{\boldmath $y$},\mbox{\boldmath $x$}')
\ln z_{1}\,,
\end{gathered}
\end{equation}
and where maximum is taken provided
\begin{equation}\label{constn1a}
\begin{gathered}
b = c + t\,, \\
cc_{1} + (1-c)c_{2} = bb_{1} + (1-b)b_{2}\,.
\end{gathered}
\end{equation}
Note that both functions $f,g$ are $\cap$--concave on all variables.

For the maximum of $f$ we have
\begin{equation}\label{PAx0}
\begin{gathered}
\max_{(\ref{constn1a})}f \leq \max_{b = c + t}f =
\max_{c}\{h(c) + h(c+t) - (2c+t)\ln z\} = \\
= h(c_{0}+t) + h(c_{0}) - (2c_{0}+t)\ln z\,,
\end{gathered}
\end{equation}
where $c_{0}(t,p)$ is defined in (\ref{Pen2}).
In fact, there is equality in (\ref{PAx0}).

To maximize the function $g$ we use the standard Lagrange
multipliers. Then for the optimal parameter values we get
$$
\begin{gathered}
c_{1} = 1 - b_{2}\,, \qquad c_{2} = 1-b_{1}\,, \qquad
b_{2} = \frac{1-(1+t)b_{1}}{1-t}\,,
\end{gathered}
$$
where $b_{1} = b_{1}(t,p_{1})$ is defined in (\ref{Pen2}). It gives
\begin{equation}\label{PBx0}
\begin{gathered}
\max_{(\ref{constn1a})}g = (1+t)h(b_{1}) +
(1-t)h\left[\frac{(1+t)b_{1}-t}{1-t}\right] -
[2+t-2(1+t)b_{1}]\ln z_{1}\,.
\end{gathered}
\end{equation}
Note that (since $z_{1} > 1$)
$$
2+t-2(1+t)b_{1} = \frac{4 + (z_{1}^{2}- 1)t^{2}}
{2 + \sqrt{4z_{1}^{2} + (z_{1}^{2}- 1)^{2}t^{2}}} > 0\,.
$$
Therefore from (\ref{defA2)}), (\ref{PAx0}) and (\ref{PBx0}) we get
\begin{equation}\label{defA20)}
\begin{gathered}
\ln {\mathbf P}({\cal A}_{1}|\mbox{\boldmath $x$}_{1}) =
-G_{2}(t,p,p_{1})m + o(n)\,,
\end{gathered}
\end{equation}
where $G_{2}(t,p,p_{1})$ is defined in (\ref{Pen2}).

Finally consider the probability
${\mathbf P}({\cal A}_{2}|\mbox{\boldmath $x$}_{1})$ from
(\ref{setA1}), (\ref{P2e}). We show that
\begin{equation}\label{A2A1}
\ln {\mathbf P}({\cal A}_{2}|\mbox{\boldmath $x$}_{1}) \leq
\ln {\mathbf P}({\cal A}_{1}|\mbox{\boldmath $x$}_{1}) + o(n)\,.
\end{equation}
For that purpose introduce the random events
$$
\begin{gathered}
{\cal C} = \left\{d(\mbox{\boldmath $x$}_{3},\mbox{\boldmath $y$})
\geq d(\mbox{\boldmath $x$}_{2},\mbox{\boldmath $y$}) + t\gamma n/2
\right\}\,,\\
{\cal D} = \left\{d(\mbox{\boldmath $x$}_{2},\mbox{\boldmath $x$}')
\leq d(\mbox{\boldmath $x$}_{3},\mbox{\boldmath $x$}') \leq
d(\mbox{\boldmath $x$}_{1},\mbox{\boldmath $x$}')\right\}
\end{gathered}
$$
and
$$
\begin{gathered}
{\cal C}_{1} = \left\{d(\mbox{\boldmath $x$}_{3},
\mbox{\boldmath $y$}) = d(\mbox{\boldmath $x$}_{2},
\mbox{\boldmath $y$}) + t\gamma n/2 + o(n)\right\}\,,\\
{\cal D}_{1} = \left\{d(\mbox{\boldmath $x$}_{2},
\mbox{\boldmath $x$}') = d(\mbox{\boldmath $x$}_{3},
\mbox{\boldmath $x$}') + o(n) = d(\mbox{\boldmath $x$}_{1},
\mbox{\boldmath $x$}') + o(n)\right\} \,.
\end{gathered}
$$
Then ${\cal A}_{2} = {\cal C} \cap {\cal D}$, and we have for any
$t \geq 0$
$$
\begin{gathered}
{\bf P}\left({\cal A}_{2}|\mbox{\boldmath $x$}_{1}\right) =
{\bf P}\left({\cal C} \cap {\cal D}|\mbox{\boldmath $x$}_{1}\right)
\sim {\bf P}\left({\cal C}_{1} \cap {\cal D}_{1}|
\mbox{\boldmath $x$}_{1}\right) \leq \\
\leq {\bf P}\left(\left\{d(\mbox{\boldmath $x$}_{3},
\mbox{\boldmath $x$}') \leq d(\mbox{\boldmath $x$}_{2},
\mbox{\boldmath $x$}')\right\} \cap {\cal C}_{1}|
\mbox{\boldmath $x$}_{1}\right)
\sim {\mathbf P}({\cal A}_{1}|\mbox{\boldmath $x$}_{1}),
\end{gathered}
$$
which proves the inequality (\ref{A2A1}).

As a result, from (\ref{P2e}), (\ref{defA20)}) and (\ref{A2A1})
we have
\begin{equation}\label{P2n}
\begin{gathered}
\frac{1}{n}\ln P_{2n} = -\frac{3\gamma}{4}G_{2}(t,p,p_{1}) + o(1)\,.
\end{gathered}
\end{equation}

For the decoding error probability $P_{\rm e}$ from (\ref{PetotaMl2}),
(\ref{Pe11a}), (\ref{Pe2}) and (\ref{P2n}) we get
\begin{equation}\label{Pe0M}
\begin{gathered}
\frac{1}{n}\ln \frac{1}{P_{\rm e}} = \max_{\gamma,t}\min\left\{
\frac{3\gamma}{4}\min\left\{G_{1}(t,p),G_{2}(t,p,p_{1})\right\},
(2-\gamma)\,E(p)\right\} = \\
= \max_{t}\frac{6\min\left\{G_{1}(t,p),G_{2}(t,p,p_{1})\right\}E(p)}
{3\min\left\{G_{1}(t,p),G_{2}(t,p,p_{1})\right\} + 4E(p)}\,,
\end{gathered}
\end{equation}
where we set
$$
\gamma = \frac{8E(p)}{3\min\left\{G_{1}(t,p),
G_{2}(t,p,p_{1})\right\} + 4E(p)}\,.
$$
The right-hand side of (\ref{Pe0M}) exceeds $E(p)$, if for some
$t$ the inequality holds
\begin{equation}\label{GGE}
3\min\left\{G_{1}(t,p),G_{2}(t,p,p_{1})\right\} > 4E(p)\,.
\end{equation}
Moreover, $t \leq 1/2-p$ (otherwise, $3G_{1}(t,p) = 4E(p)$).
Since $G_{2}(t,p,p_{1})$ monotonically increases in $t$, in order
to have the inequality (\ref{GGE}) fulfilled, we need to have
$3G_{2}(1/2-p,p,p_{1}) > 4E(p)$. Therefore introduce
the function $p_{0}(p)$ as the unique root of the equation
(\ref{defp0}). Then for any $p_{1} < p_{0}(p)$ and some $t < 1/2-p$
the inequality (\ref{GGE}) is fulfilled, and therefore the
right-hand side of (\ref{Pe0M}) exceeds $E(p)$. As a result, from
(\ref{Pe0M}) we get the formula (\ref{Pe00}), which proves the
theorem.  $\qquad \Box$

\medskip

\newpage
\begin{center}
{\large\bf \S\,4.  Channel with active feedback. Example}
\end{center}

Using of coding in the feedback channel enlarges transmission
possibilities. As an example, we consider the simplest of such
transmission methods, proposed by G.A. Kabatyansky. The
transmitter and the receiver will send information by turns.

We set some numbers $\gamma, \gamma_{1} > 0$, such that
$\gamma + \gamma_{1} < 1$, and divide the total transmission period
$[0,n]$ on intervals
$[0,\gamma n], (\gamma n, (\gamma + \gamma_{1})n]$ and
$((\gamma + \gamma_{1})n,n]$. We call those intervals phases
I, II and III, respectively.

The transmitter will send information on phases I and III, while
the receiver will send information only on phase II. On phase I of
length $\gamma n$ we use ``almost'' a simplex code. After phase I,
based on the received block $\mbox{\boldmath $y$}$, the receiver
selects two most probable messages. Then, during the phase II of
length $\gamma_{1} n$, it informs the transmitter on those two
messages. On phase III, the transmitter uses two opposite codewords
of length $(1-\gamma - \gamma_{1}) n$ to help the receiver to decide
between those two most probable messages.

A decoding error occurs in the following three cases:

1) After phase I the true message is not among two most probable
messages. We denote that probability $P_{1}$.

2) After phase I the true message is among two most probable, but
on phase II the decoding error occurs on the transmitter. We denote
that probability $P_{2}$.

3) After phase II the transmitter identified correctly two most
probable messages (and the true message is among them), but after
phase III the true message is not the most probable one among two
possible messages. We denote that probability $P_{3}$.

Then for the decoding error probability $P_{\rm e}$ we have
\begin{equation}\label{Pecodl}
P_{\rm e} \leq P_{1} + P_{2} + P_{3}\,.
\end{equation}
Similarly to \S\,3, for the values $P_{1}, P_{2}, P_{3}$ in the
right-hand side of (\ref{Pecodl}) we have (as $n \to \infty$)
\begin{equation}\label{Pecod2}
\begin{gathered}
\frac{1}{n}\ln \frac{1}{P_{1}} = \frac{3}{4}\gamma F(p) + o(1)\,,\\
\frac{1}{n}\ln \frac{1}{P_{2}} = \gamma_{1}E(p_{1}) + o(1)\,, \\
\frac{1}{n}\ln P_{3} = (2-\gamma - 2\gamma_{1})E(p) + o(1)\,.
\end{gathered}
\end{equation}
We choose parameters $\gamma,\gamma_{1}$ such that the values
$P_{1}, P_{2}, P_{3}$ become equal, i.e. we set
$$
\gamma_{1} = \frac{3\gamma F(p)}{4E(p_{1})}\,, \qquad
\gamma = \frac{8E(p)}{3F(p) + 4E(p) + 6F(p)E(p)/E(p_{1})}\,.
$$
Then we get

\medskip

{P r o p o s i t i o n}. {\it For the decoding error probability
$P_{\rm e}$ of the transmission method described the relation holds}
\begin{equation}\label{Pecod3}
\begin{gathered}
\frac{1}{n}\ln \frac{1}{P_{\rm e}} \geq
\frac{E(p)}{1/2 + 2E(p)/(3F(p)) + E(p)/E(p_{1})} + o(1)\,.
\end{gathered}
\end{equation}

\medskip

Since $E(p)/F(p) \to 3/4,\,p \to 0$, such transmission method,
essentially, does not improve $E(p)$ for small $p$
(and any $p_{1}$).

But if $p=(1-\varepsilon)/2,\,\varepsilon \to 0$, then
$E(p)/F(p) \to 9/16,\,p \to 1/2$, and (\ref{Pecod3}) takes the form
\begin{equation}\label{Pecod3a}
\begin{gathered}
\frac{1}{n}\ln \frac{1}{P_{\rm e}} \geq
\frac{E(p)}{7/8 + E(p)/E(p_{1})} + o(1)\,.
\end{gathered}
\end{equation}
In that case, such transmission method improves $E(p)$, if
$E(p_{1}) > 8E(p)$. In particular, if
$p_{1} =(1-\varepsilon_{1})/2,\,\varepsilon_{1} \to 0$, then
$E(p)/E(p_{1}) \approx \varepsilon^{2}/\varepsilon_{1}^{2}$.
Therefore the right-hand side of (\ref{Pecod3a}) is better than
$E(p)$, if $\varepsilon_{1} > \varepsilon \sqrt{8}$. It is better
than the relation (\ref{Pe00}) (where it is demanded
$p_{1} < 0.067$).

\medskip

\begin{center} {\large Acknowledgment } \end{center}

The authors wish to thank the University of Tokyo for supporting
this joint research.

\newpage

\hfill {\large\sl APPENDIX}

\medskip

{P r o o f \, o f \, l e m m a}. Since the part 2) follows from
the part 1), it is sufficient to prove the part 1). To simplify
formulas we assume that $d_{12} = d_{13} = d_{23} = 2m/3$ (i.e. that
$\{\mbox{\boldmath $x$}_{i}\}$ is a simplex code). Such codewords
$\mbox{\boldmath $x$}_{1},\mbox{\boldmath $x$}_{2},
\mbox{\boldmath $x$}_{3}$ are shown in Fig. 4, where
$a, b, c$ denote the fractions of $1$'s in the corresponding parts
of the received block $\mbox{\boldmath $y$}$. Since
\begin{equation}\label{dist1}
\begin{gathered}
d_{1} = d(\mbox{\boldmath $x$}_{1},\mbox{\boldmath $y$}) =
(a+b+c)m/3\,, \\
d_{2} = d(\mbox{\boldmath $x$}_{2},\mbox{\boldmath $y$}) =
(2+c-a-b)m/3\,, \\
d_{3} = d(\mbox{\boldmath $x$}_{3},\mbox{\boldmath $y$}) =
(2+b-a-c)m/3\,,
\end{gathered}
\end{equation}
for the corresponding random events we have
$$
\begin{gathered}
\{d_{2} = d_{1} +  2tm/3\} \Leftrightarrow \{a+b=1-t \}, \\
\{d_{3} = d_{1} + 2t_{1}m/3\} \Leftrightarrow \{a+c=1-t_{1} \}.
\end{gathered}
$$
Therefore
$$
\begin{gathered}
P_{1}(t,t_{1}) \sim q^{m} \max_{\substack {a+b = 1-t \\
a+c = 1-t_{1}}}
\left\{\binom{m/3}{am/3}\binom{m/3}{bm/3}\binom{m/3}{cm/3}
z^{-(a+b+c)m/3}\right\},
\end{gathered}
$$
and then
$$
\begin{gathered}
\frac{3}{m}\ln P_{1}(t,t_{1}) = \ln(p^{2}q) +
\max_{a} f(a,t,t_{1}) + o(1)\,,
\end{gathered}
$$
where
$$
\begin{gathered}
f(a,t,t_{1}) = h(a)+ h(a+t) + h(a+t_{1}) + (a+t_{1}+t)\ln z \,, \\
f'_{a} = \ln\frac{1-a}{a} + \ln\frac{1-a-t}{a+t} +
\ln\frac{1-a-t_{1}}{a+t_{1}} + \ln z\,, \\
f'_{t} =  \ln\frac{1-a -t}{a+t} + \ln z\,, \qquad
f'_{t_{1}} = \ln\frac{1-a-t_{1}}{a+t_{1}} + \ln z\,.
\end{gathered}
$$
The function $f(a,t,t_{1})$ is $\cap$--concave on all arguments.
Therefore, the function $\max\limits_{a}f(a,t,t_{1})$ (and similar
ones) is also $\cap$--concave on all arguments. In particular,
$\max\limits_{a,t,t_{1}}f(a,t,t_{1})$ is attained for
$a=p,\,t=t_{1}=1-2p$. Similarly, $\max\limits_{a,t_{1}}f(a,t,t_{1})$
is attained for $a=(1-t)/2,\,t_{1}=(1-2p+t)/2$. Then we get the part
1) of the lemma.  $\qquad \Box$


\vspace{5mm}

\begin{flushleft}
{\small {\it Burnashev Marat Valievich} \\
Institute for Information Transmission Problems RAS \\
 {\tt burn@iitp.ru}} \\
 {\small {\it Yamamoto Hirosuke} \\
The University of Tokyo, Japan \\
 {\tt hirosuke@ieee.org}}
\end{flushleft}%

\newpage

\linethickness{0.2mm}

\begin{picture}(110,130) (-30,100)

\put(160,110){\circle*{3}}
\put(200,90){\circle*{3}}

\put(160,120){\line(0,1){60}}
\put(160,120){\line(-5,-3){80}}
\put(160,120){\line(5,-3){80}}


\put(147,103){$A$}
\put(200,80){$B$}
\put(155,58){${\cal D}_{1}$}
\put(215,138){${\cal D}_{2}$}
\put(95,138){${\cal D}_{3}$}

\put(160,110){\vector(0,-1){20}}
\put(200,90){\vector(-1,-1){15}}

\end{picture}

\medskip

\vspace{2.0cm}

\begin{center}  Fig 1. \ Decoding regions
${\cal D}_{1},{\cal D}_{2},{\cal D}_{3}$ and directions of output
drives
\end{center}

\newpage

\includegraphics[width=0.9\hsize,height=0.8\hsize]{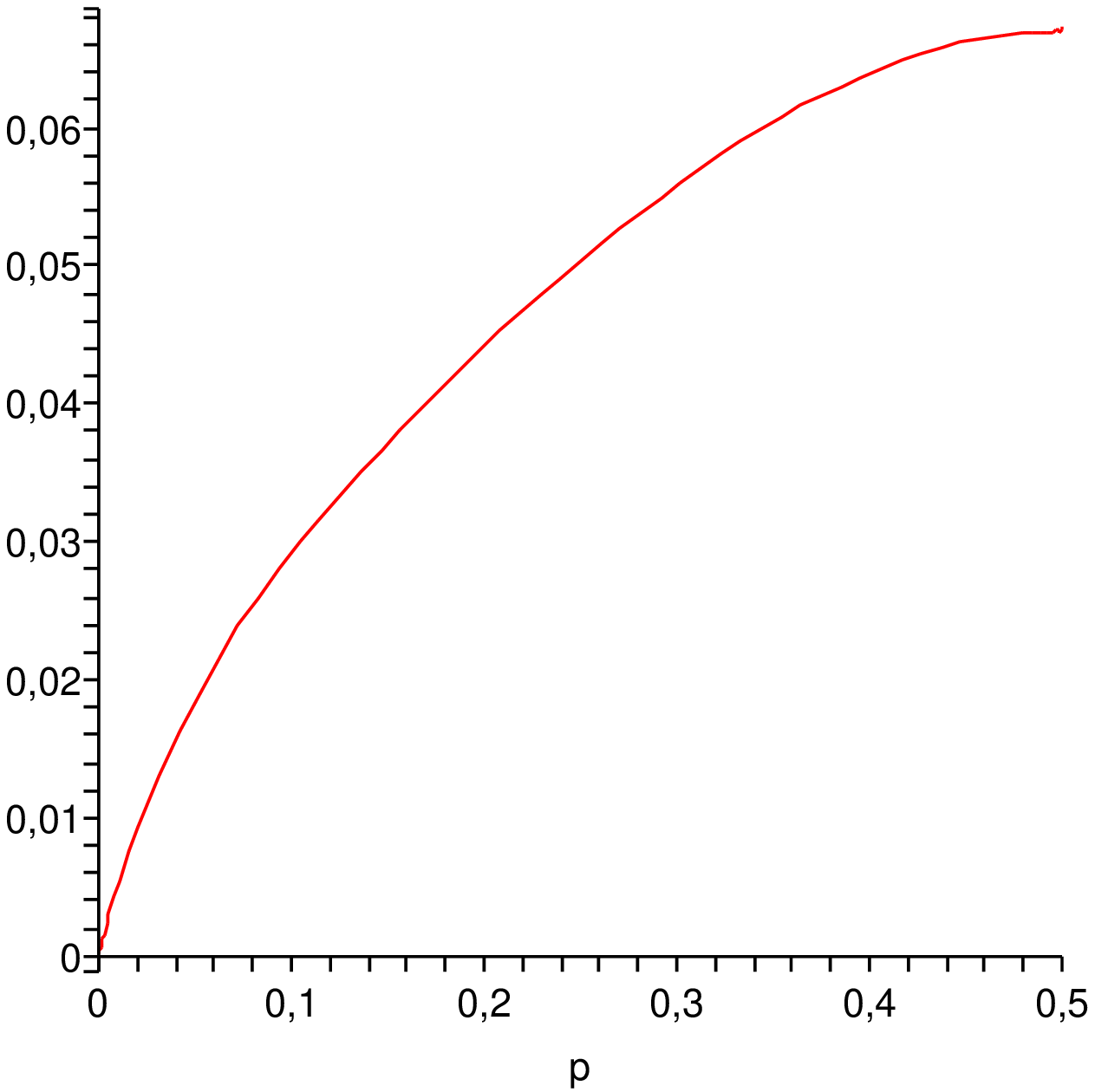}

\begin{center} { Fig. 2. Plot of the function $p_{0}(p)$}
\end{center}

\newpage

\linethickness{0.3mm}

\begin{picture}(150,50) (-90,100)

\put(10,60){\line(0,1){100}}
\put(40,60){\line(0,1){100}}
\put(80,60){\line(0,1){100}}
\put(110,60){\line(0,1){100}}
\put(150,60){\line(0,1){100}}
\put(180,60){\line(0,1){100}}
\put(220,60){\line(0,1){100}}

\put(-15,145){$\mbox{\boldmath $x$}_{1}$}
\put(-15,125){$\mbox{\boldmath $x$}_{2}$}
\put(-15,105){$\mbox{\boldmath $x$}_{3}$}
\put(-15,85){$\mbox{\boldmath $y$}$}
\put(-15,65){$\mbox{\boldmath $x$}'$}

\put(20,145){$00$}
\put(55,145){$00$}
\put(90,145){$00$}
\put(125,145){$00$}
\put(160,145){$00$}
\put(195,145){$00$}

\put(20,125){$11$}
\put(55,125){$11$}
\put(90,125){$11$}
\put(125,125){$11$}
\put(160,125){$00$}
\put(195,125){$00$}

\put(24,65){$a_{1}$}
\put(56,65){$a_{2}$}
\put(93,65){$b_{1}$}
\put(126,65){$b_{2}$}
\put(162,65){$c_{1}$}
\put(196,65){$c_{2}$}

\put(20,105){$11$}
\put(55,105){$11$}
\put(90,105){$00$}
\put(125,105){$00$}
\put(160,105){$11$}
\put(195,105){$11$}

\put(20,85){$11$}
\put(55,85){$00$}
\put(90,85){$11$}
\put(125,85){$00$}
\put(160,85){$11$}
\put(195,85){$00$}

\put(10,45){$0$}
\put(74,45){$m/3$}
\put(140,45){$2m/3$}
\put(215,45){$m$}

\put(36,165){$a$}
\put(106,165){$b$}
\put(176,165){$c$}

\end{picture}

\vspace{2.0cm}

\begin{center}
Fig 3. \ Blocks
$\mbox{\boldmath $x$}_{1},\mbox{\boldmath $x$}_{2},
\mbox{\boldmath $x$}_{3},\mbox{\boldmath $y$},\mbox{\boldmath $x$}'$
\end{center}

\newpage

\linethickness{0.3mm}

\begin{picture}(160,20) (-100,100)

\put(80,60){\line(0,1){80}}
\put(110,60){\line(0,1){80}}
\put(140,60){\line(0,1){80}}
\put(170,60){\line(0,1){80}}

\put(55,125){$\mbox{\boldmath $x$}_{1}$}
\put(55,105){$\mbox{\boldmath $x$}_{2}$}
\put(55,85){$\mbox{\boldmath $x$}_{3}$}
\put(55,65){$\mbox{\boldmath $y$}$}

\put(90,125){$00$}
\put(93,65){$a$}
\put(120,125){$00$}
\put(123,65){$b$}
\put(150,125){$00$}
\put(153,65){$c$}

\put(90,105){$11$}
\put(120,105){$11$}
\put(150,105){$00$}

\put(90,85){$11$}
\put(120,85){$00$}
\put(150,85){$11$}

\put(80,45){$0$}
\put(98,45){$m/3$}
\put(128,45){$2m/3$}
\put(166,45){$m$}

\end{picture}

\vspace{2.0cm}

\begin{center}
Fig 4. \ Blocks $\mbox{\boldmath $x$}_{1},\mbox{\boldmath $x$}_{2},
\mbox{\boldmath $x$}_{3},\mbox{\boldmath $y$}$
\end{center}


\medskip

\newpage

\begin{center} {\large REFERENCES} \end{center}
\begin{enumerate}
\bibitem{Shannon56}
{\it Shannon C. E.} The Zero Error Capacity of a Noisy Channel //
IRE Trans. Inform. Theory. 1956. V. 2. ? 3. P. 8--19.
\bibitem{Dob}
{\it Dobrushin R. L.} Asymptotic bounds on error probability for
message transmission in a memoryless channel with feedback //
Probl. Kibern. No. 8. M.: Fizmatgiz, 1962. P. 161--168.
\bibitem{Hors1}
{\it Horstein M.} Sequential Decoding Using Noiseless Feedback //
IEEE Trans. Inform. Theory. 1963. V. 9. ? 3. P. 136--143.
\bibitem{Ber1}
{\it Berlekamp E. R.}, Block Coding with Noiseless Feedback,  Ph. D.
Thesis, MIT, Dept. Electrical Enginering, 1964.
\bibitem{Bur1}
{\it Burnashev M. V.} Data transmission over a discrete channel with
feedback: Random transmission time // Problems of Inform. Transm.
1976. V. 12, ? 4. P. 10--30.
\bibitem{Bur2}
{\it Burnashev M. V.} On a Reliability Function of Binary Symmetric
Channel with \\ Feedback // Problems of Inform. Transm.
1988. V. 24, ? 1. P. 3--10.
\bibitem{Pin1}
{\it Pinsker M. S.} The probability of error in block transmission
in a memoryless Gaussian channel with feedback // Problems of
Inform. Transm. 1968. V. 4, ? 4. P. 3--19.
\bibitem{SchalKai}
{\it Schalkwijk J. P. M., Kailath T.} A Coding Scheme for Additive
Noise Channels with Feedback - I: No Bandwidth Constraint // IEEE
Trans. Inform. Theory. 1966. V. 12. ? 2. P. 172--182.
\bibitem{Tchamtel1}
{\it Tchamkerten  A., Telatar E.} Variable Length Coding over an
Unknown Channel // IEEE Trans. Inform. Theory. 2006. V. 52. ? 5.
P. 2126--2145.
\bibitem{YamIt}
{\it Yamamoto H., Itoh R.} Asymptotic Performance of a Modified
Schalkwijk--Barron \\ Scheme for Channels with Noiseless Feedback //
IEEE Trans. Inform. Theory. 1979. V. 25.  ? 6. P. 729--733.
\bibitem{DrapSah1}
{\it Draper S. C., Sahai A.} Noisy Feedback Improves Communication
Reliability // Proc. IEEE International Symposium on
Information Theory. Seattle, WA, July 2006, P. 69--73.
\bibitem{KimLapW}
{\it Kim Y.-H., Lapidoth A., Weissman T.} The Gaussian Channel with
Noisy Feedback //  Proc. IEEE International Symposium on
Information Theory, Nice,  France, June 2007, P. 1416--1420.
\bibitem{BurYam1}
{\it Burnashev M. V., Yamamoto H.} On BSC, Noisy Feedback and Three
Messages //  Proc. IEEE Int. Sympos. on Information Theory.
Toronto,  Canada. July, 2008. P. 886--889.
\end{enumerate}
\end{document}